\documentstyle[prd,preprint,aps]{revtex}

\begin{document}

\reversemarginpar

\title{Schwarzschild black hole as a grand canonical ensemble}
\author{Gilad Gour\thanks{E-mail:~gour@cc.huji.ac.il}} 
\address{Racah Institute of Physics, Hebrew University of Jerusalem,\\ 
Givat Ram, Jerusalem~91904, ISRAEL.}

\maketitle

\begin{abstract}

For long black holes have been considered as endowed with a definite
temperature. Yet when the Schwarzschild black hole is treated as a canonical 
ensemble three problems arise: incompatibility with the
Hawking radiation, divergence of the partition function, and a formally
negative mean-square fluctuation of the energy. We solve all three 
problems by considering the Schwarzschild black hole as a grand canonical
ensemble, with the Hamiltonian (the ADM mass) and the horizon surface 
area, separately, as observable parameters. 
The horizon area simulates the number of particles in statistical 
mechanics since its spectrum is here assumed to be discrete and equally spaced. 
We obtain a logarithmic correction to the Bekenstein-Hawking entropy 
and a Gaussian type distribution for the energy levels. 
  
\end{abstract}

\pacs{PACS numbers:~}

Black holes (BHs) are conventionally regarded as thermodynamic 
systems~\cite{Bek73,Wald94}. But there exist problems with the 
description of a black hole as a canonical ensemble~\cite{Hawk76}.
For example, because a BH has negative specific heat, energy 
fluctuations calculated in canonical ensemble have formally negative
variance. The issue of how to give a consistent thermodynamic description
of a BH has received renewed urgency with the understanding that the
mass spectrum of a BH may be discrete and highly degenerate. 

In the early seventies, Bekenstein pointed out that the horizon area of 
a non-extermal BH behaves as a classical adiabatic 
invariant~\cite{BekLectures,Bek74}. Using Ehrenfest's
principle~\cite{Eren} that any classical adiabatic invariant corresponds to a
quantum entity with discrete spectrum, Bekenstein conjectured that the
spectrum of the horizon area of a (non--extremal) BH should be quantized.
He proposed a uniformly spaced spectrum which has later been considered
by many authors (see the list in Kastrup~\cite{Kast97}). 
Today the idea of a discrete eigenvalue spectrum for the horizon area is also
supported by the work of Ashtekhar and others~\cite{AshKra}, and recently a 
uniformly spaced spectrum (for large quantum numbers) has been established by Bojowald
and Kastrup~\cite{BoKa99} within the framework of loop quantum gravity.

Recently, the Schwarzschild BH with the uniformly spaced area spectrum 
was treated as a microcanonical ensemble by Scharf~\cite{Scharf97}
(a microcanonical ensemble of a gas of neutral and charged BHs has been 
considered earlier by Harms and Leblanc~\cite{HarLe92}) and as a quantum 
canonical ensemble by Kastrup~\cite{Kast97} and by M\"{a}kel\"{a}
and Repo~\cite{MakRep98}. According to the ``no hair'' 
theorems~\cite{BekLectures,NoHair}, an observer outside 
a Schwarzschild black hole is able to 
measure only its mass. Hence, all authors cited above have used the mass (energy) 
of the BH as the sole variable characterizing the system. The main new idea of 
the present paper is that a Schwarzschild BH formed by gravitational collapse 
should rather be considered as a grand canonical ensemble. 
The additional thermodynamic parameter is the horizon surface area of the BH. 

In the present paper we shall assume that the area 
spectrum $A_{n}$ of the Schwarzschild BH is given by:
\begin{equation}
A_{n}=a_{0}n\;\;\;n=0,1,2,...
\label{arlev}
\end{equation}
where $a_{0}$ is a constant of the order of the Planck area. Recently, Kastrup
and others~\cite{KasOth} have shown that Eq.~(\ref{arlev}) holds true also when
the space-time dimensions is taken to be $D>4$.  

The area levels $A_{n}$ are expected to be degenerate. Denoting the
degeneracy by $g(n)$ and identifying $\ln g(n)$ with the BH entropy
\begin{equation}
S=A/(4{\cal L}^{2}_{p})
\label{int}
\end{equation} 
(${\cal L}_{P}$ and ${\cal M}_P$ will here denote the Planck length and mass, respectively),
Bekenstein and Mukhanov~\cite{BekMu95} found that
$a_{0}=4{\cal L}_{p}^{2}\ln k$,
or equivalently $g(n)=k^{n}$ where $k=2,3,4,\dots$ They adopted $k=2$ for
simplicity; recently Hod~\cite{Hod2}  has argued in favor of
the choice $k=3$.
 
In the quantum mechanical picture, if an observer at infinity 
makes an attempt to determine the mass of a black 
hole, his accuracy is limited by the time-energy uncertainty 
relation as well as by the systematic 
decrease of the mass of the BH~\cite{Hawk75}. 
Thus for a black 
hole formed by a gravitational collapse, it is impossible, even conceptually, 
for an observer to know exactly the mass of the black hole. 
However, if the black hole is in a static
state (eternal black hole), the observer, in principle, would be able to determine 
its mass and then the microcanonical approach would be appropriate. 
   
In the canonical ensemble approach the observer knows only the average value of the mass of 
the black hole at a given time. In this ensemble the partition function $Z$ formally 
diverges because the degeneracy factor $g(n)$ rises very fast with $n$. 
Kastrup~\cite{Kast97} proposes to resolve this problem by defining $Z$ by analytic 
continuation, a procedure which is difficult to understand in physical terms. 
M\"{a}kel\"{a} and Repo~\cite{MakRep98} avoided the problem by studying, not
the partition function of the whole spacetime itself, but instead the partition 
function of the radiation emitted by the BH.
We shall see shortly that if we treat the BH as a grand canonical 
ensemble, $Z$ is no longer divergent! Before describing this new
approach we shall first point out another problem of the canonical approach. 

Treating the BH as a canonical ensemble implies that it is in a thermal
equilibrium with a surrounding thermal bath. 
Thus the BH mass remains constant because the density matrix of the canonical 
ensemble is constant in time; it commutes with the Hamiltonian 
(the ADM mass operator). However, the mass of an isolated BH formed by 
a gravitational collapse decreases in time because of losses to 
Hawking radiation~\cite{Hawk75}, so a description of it via a thermal ensemble
seems inappropriate. We propose to solve both problems by abandoning the 
canonical approach in favor of the grand canonical one.     

If the observer is interested in determining the horizon area of the black hole
he is limited, 
apart from the time-energy uncertainty relation, by some kind of  
area-phase uncertainty relation~\cite{Gour}:
\begin{equation}
\Delta{\bf A}\Delta{\bf\phi}\geq\frac{1}{2}a_{0}.
\label{aphu}
\end{equation}
Here ${\bf\phi}$ is the canonical conjugate to the area (number) operator.
This is the first clue suggesting
to treat the black hole as a grand canonical ensemble: the observer only
knows the average value of the horizon area (by Eq.~(\ref{aphu}) the area operator
is a number operator, and is thus analogous to the number of particles 
in a grand canonical system). 
We shall discuss now further physical grounds for adopting the 
grand canonical approach for the description of Schwarzschild BHs.

Classically, the ADM mass and the horizon surface area are related
by
\begin{equation}
A=\frac{16\pi G^{2}}{c^{2}}M^{2}.
\label{clasrel}
\end{equation} 
Hence, the horizon area was never before considered as a new 
parameter. However, the mass (energy) and the horizon area (topology feature) 
of a BH  describe two {\it different} properties.

Consider an observer who is not aware of the classical relation in 
Eq.~(\ref{clasrel}), and is interested in the BH properties. It is 
clear that he would use completely different techniques (and apparatus) 
to measure the mass and the horizon area of the BH. The mass may be
measured asymptotically at infinity whereas the horizon surface area
should be measured locally. Thus, these two parameters describe operationally 
distinguishable features of the BH.

In the quantum mechanical picture the distinction between these two
parameters becomes prominent. Because of the Hawking phenomenon, the
horizon area decreases in time and hence the area operator and the Hamiltonian
(the ADM mass operator) do not commute. Thus, the relation in 
Eq.~(\ref{clasrel}) does not hold true for operators. In~\cite{Gour} we have found 
the form of the Hamiltonian and have shown that in the classical limit, where
the HR is negligible, Eq.~(\ref{clasrel}) is satisfied also for the operators.

Generally, in grand canonical ensembles the number of particles and the energy
of the system are taken to be the two observable parameters since neither is 
constant and each describes a {\it different} property
of the system (even though they are ultimately related by some formula). 
Since the ADM mass
and the horizon area of a Schwarzschild BH have the same relation, we conclude
that a Schwarzschild BH should be considered as a grand canonical ensemble!
 
Assuming the black hole is described by some density operator ${\bf\rho}$, we shall
maximize the following quantity (entropy):
\begin{equation}
Q=-{\rm Tr}\left({\bf \rho}\ln {\bf \rho}\right)-\mu'\langle {\bf A}\rangle
-b\langle {\bf H}\rangle
\label{mq}
\end{equation}
where $\mu'$ and $b$ are the Lagrangian multipliers (the physical meaning of these
parameters will be discussed later) and ${\bf H}$ is the Hamiltonian
(ADM mass) operator (boldface is used everywhere to denote operators). The trace
in Eq.~(\ref{mq}) may be taken with respect to the area eigenstates; this makes it easy
to take the degeneracy factor $g(n)$ into account. 
The extremum for $Q$ under the conditions that 
${\rm Tr}\left({\bf \rho} {\bf A}\right)\equiv\langle {\bf A}\rangle$ and 
${\rm Tr}\left({\bf \rho} {\bf H}\right)\equiv\langle {\bf H}\rangle$ 
with $\langle {\bf A}\rangle$ and $\langle {\bf H}\rangle$ known
is attained by
\begin{equation}
{\bf\rho}=\frac{1}{Z}\exp\left(-\mu' {\bf A}-b{\bf H}\right),
\label{densop}
\end{equation}
where the partition function $Z$ is defined by
\begin{equation}
Z={\rm Tr}\left(\exp(-\mu' {\bf A}-b{\bf H})\right).
\label{part}
\end{equation}

Because of the Hawking radiation, the Hamiltonian operator cannot commute with
the horizon area operator, because the last operator is not constant in time.
Thus ${\bf\rho}$ does not commute with ${\bf H}$. 
Hence, we conclude that by choosing the 
appropriate Hamiltonian operator~\cite{Gour} for the Schwarzschild BH, the 
density matrix in Eq.~(\ref{densop}) would be compatible with BH evolution
in the wake of Hawking radiation. 

We have shown in ref.~\cite{Gour} that the Hamiltonian (the ADM mass) can be written as 
\begin{equation}
{\bf H}={\bf M}+{\bf V}
\label{hamil}
\end{equation}
where ${\bf M}\equiv\sqrt{c^{4}{\bf A}/16\pi G^{2}}$ 
is the mass operator (according to the classical limit) 
and ${\bf V}$ is a coupling between the horizon area and its 
canonical conjugate, the phase of the BH. However, as we have pointed out, 
the interaction term approaches zero as 
$\langle {\bf M}\rangle\rightarrow\infty$ according to 
$\langle {\bf V}\rangle\sim\ {\cal M}_{p}^{2} c^{2} /\langle {\bf M}\rangle$.
Hence, for BHs not near the Planck scales we may neglect the effect of Hawking radiation 
and assume that ${\bf H}\approx {\bf M}$. Thus, the partition function given 
in Eq.~(\ref{part}) can be approximated by  
\begin{equation}
Z=\sum_{n=0}^{\infty} k^{n}\exp\left(-\mu' a_{0}n-bm_{0}\sqrt{n}\right)
=\sum_{n=0}^{\infty}\exp\left(-(\mu' a_{0}-\ln k)n-bm_{0}\sqrt{n}\right),
\label{compari}
\end{equation}
where $m_{0}\equiv\sqrt{c^{4}a_{0}/16\pi G^{2}}$ 
is of the order of the Planck mass.
Let us define the dimensionless coefficients  $\alpha\equiv a_{0}\mu' -\ln k$
and $\chi\equiv bm_{0}$. We note that the partition function converges for $\alpha >0$;
we thus assume $\alpha >0$ and that $\alpha$ is crudely of order unity.
The probability to find the system in the $n$th area state can be written as
\begin{equation}
P_{n}=\frac{1}{Z}\exp\left(-\alpha n-\chi\sqrt{n}\right)
\label{pon}
\end{equation}
where $\alpha$ and $\chi$ must be reexpressed in terms of $\langle {\bf A}\rangle$ and
$\langle {\bf M}\rangle$.

The relation between $\langle {\bf A}\rangle$
and $\langle {\bf M}\rangle$ is not exactly $\langle {\bf A}\rangle$=$
16\pi\langle {\bf M}\rangle^{2}G^{2}/c^{4}$ since the expectation values are not
taken with respect to a pure state with sharp mass. What would it be for
a BH formed by gravitational collapse?
In order to answer, let us first introduce a new parameter $n_{0}$,
the number $n$ that maximizes $P_{n}$. It is clear that around this
value the distribution is symmetric, that is
\begin{equation}
P_{n_{0}+h}\approx P_{n_{0}-h} 
\label{np}
\end{equation}
for $h\ll n_{0}$. Comparing Eq.~(\ref{np}) with Eq.~(\ref{pon}) we find that
\begin{equation} 
\chi=-2\sqrt{n_{0}}\alpha.
\label{probl}
\end{equation}
Eq.~(\ref{probl}) raises a serious problem regarding the physical meaning 
of the parameters $\mu'$ and $b$. A comparison of the partition 
function~(\ref{compari}) with the partition function
\begin{equation}
Z = \sum_n g(E_n)\exp\left(\beta(\mu N_n - E_n)\right)
\end{equation}
written for a general grand canonical ensemble with energy levels $E_n$,
particle number $N_n$, degeneracy $g(E_n)$, chemical potential $\mu$ and
inverse temperature $\beta$ reveals that the parameter $\mu'$ in our analysis
represents the negative of the chemical potential divided by the temperature, 
and parameter $b$ the inverse temperature of the BH. Now, Eq.~(\ref{probl})
implies that $\chi$ is negative because $\alpha>0$ and thus also $b=\chi/m_{0}$
is negative. Does this imply that either the BH's mass or its temperature 
are negative?

In the view of M\"{a}kel\"{a} and Repo~\cite{MakRep98}, 
one can solve the problem by assuming $m_{0}$
is negative. In that case, $b$ is positive (and thus the temperature too) and
$M_{n}$ increases when $n$ decreases. In other words, $M_{n}$ becomes greater
when the BH becomes smaller. The meaning of $M_{n}$ changes: it is not the mass of 
the BH but, the {\it energy of the BH radiation} (with an appropriate choice 
of energy zero). 
Here we suggest another solution to the problem which also saves the 
positivity of both the mass and the temperature of the BH, even though $b$ is 
negative. 

Since the horizon splits the space into two parts, we cannot immediately compare
our analysis with the one in general grand canonical ensemble. According to 
the ``no hair" theorems, the degrees of freedom in the interior region are 
not accessible to an observer at infinity. This affects the independence
of the horizon surface area and the mass of the BH. Independence of $A$ on 
$M$ implies, for example, that it is possible to change $A$ (at least slightly) while
keeping $M$ constant. Hence, at infinity 
one cannot observe the independence of $A$ and $M$ due to the ``no hair" theorems.

The partition function that maximizes the entropy of the whole BH spacetime is 
given by Eq.~(\ref{compari}). Now, in order to associate with the BH  
a temperature $T_{BH}=\hbar/8\pi M$, we must restrict ourselves to the exterior 
region of the BH; there is no meaning to temperature ``inside the BH".
How is this restriction implemented in our scheme?

According to the above arguments we have to make the transition from two independent 
parameters $A$ and $M$ to one parameter. Normally, in the usual grand canonical ensemble,
the parameter $b$ would be given by
\begin{equation}
b=\left(\frac{\partial S}{\partial\langle 
{\bf M}\rangle}\right)_{\langle {\bf A}\rangle}.
\label{demotem}
\end{equation}
Note that this derivative would be taken with respect to $\langle {\bf M}\rangle$ while
keeping $\langle {\bf A}\rangle$ constant, and would be negative. 
But when we are restricted to the exterior 
region, there is only one parameter, say $\langle {\bf M}\rangle$, and therefore the
inverse temperature is defined by (the entropy is derived in Eq.~(\ref{entrop}))
\begin{equation}
\beta\equiv\left(\frac{dS}{d\langle {\bf M}\rangle}\right)
=\frac{8\pi\langle {\bf M}\rangle}{\hbar}+{\rm O}\left(\frac{1}{\langle {\bf M}\rangle}\right)
\label{asd}
\end{equation}
which is positive and equal to the inverse of Hawking's temperature. 
Because the derivative in~Eq.~(\ref{asd}) is the total
derivative, it is distinct from $b$, and can be positive. 
No negative temperature is necessary.  
 
Substituting Eq.~(\ref{probl}) back in Eq.~(\ref{pon})
we find in the limit $n_{0}\rightarrow\infty$
\begin{equation}
P_{n}=\frac{1}{Z}\exp \left(-\alpha(n-2\sqrt{n_{0}n}\right))
\approx\left(\frac{\alpha}{4\pi n_{0}}\right)
^{\frac{1}{2}}\exp\left(-\frac{\alpha}{4n_{0}}(n-n_{0})^{2}\right).
\label{twe}
\end{equation}
Hence, we obtain a Gaussian distribution with a variance 
$\sigma_{A}=\sqrt{2n_{0}/\alpha}$. 
Note that $n_{0}$ is approximately the average of ${\bf N}\equiv {\bf A}/a_{0}$. 
Thus, as is typical of many-particle statistical systems, as 
$\langle {\bf N}\rangle\rightarrow\infty$, the absolute fluctuations become large,
but the relative fluctuations approach zero. 

The entropy may now be expressed as
\begin{equation}
S=-{\rm Tr}\left({\bf \rho}\ln {\bf \rho}\right)=\mu'\langle {\bf A}\rangle
+b\langle {\bf M}\rangle+\ln Z
\label{entro}
\end{equation}
where for large $n_{0}$ the partition function~\ref{compari} is given up to 
a very good approximation by
\begin{equation}
Z=\exp(\alpha n_{0})\left(\sqrt{\frac{4\pi n_{0}}{\alpha}}
+{\rm O}(1/n_{0}^{1/2})\right).
\label{parti}
\end{equation}
Calculating $\langle {\bf A}\rangle$, $\langle {\bf M}\rangle$ with~(\ref{twe}) 
and  taking into account the first order corrections we find 
\begin{eqnarray} 
\langle {\bf A}\rangle & = & a_{0}n_{0}+\frac{3a_{0}}{2\alpha}
+{\rm O}(1/n_{0})\nonumber\\
\langle {\bf M}\rangle & = & m_{0}\sqrt{n_{0}}+\frac{m_{0}}{2\alpha\sqrt{n_{0}}}
+{\rm O}(1/n_{0}^{3/2}).
\label{avma}
\end{eqnarray}
Substituting all these in Eq.~(\ref{entro}) we finally obtain
\begin{equation}
S=\frac{1}{4{\cal L}_{p}^{2}}\langle {\bf A}\rangle
+\frac{1}{2}\ln\left(\frac{\langle {\bf A}\rangle}{{\cal L}_{p}^{2}}\right)
+\frac{1}{2}\ln\left(\frac{4\pi}{\alpha}\right)
+\frac{3(\ln k-1)}{2\alpha}+\frac{1}{2}.
\label{entrop}
\end{equation}
Note that the main contribution to
the entropy is given by $\langle {\bf A}\rangle/4{\cal L}_{p}^{2}$ as 
was to be expected. The logarithmic correction to the entropy is exactly the same as
M\"{a}kel\"{a} and Repo obtained for the emitted radiation~\cite{MakRep98}
and as Kastrup obtained from his analytic continuation approach~\cite{Kast97}. 
However, our grand canonical approach does not suffer from a divergent partition 
function. Furthermore, using Eq.~(\ref{avma}) to determine
the fluctuation in the mass we find
\begin{equation}
\sigma_{M}^{2}\equiv\langle {\bf M}^{2}\rangle-\langle {\bf M}\rangle ^{2}
=\frac{\hbar\ln k}{8\pi\alpha}.
\end{equation}
Thus, the mean square fluctuations of the energy is {\it positive}
even while the specific heat is {\it negative}.

In summary the grand canonical approach solves three problems which arise when using the
canonical approach with the area spectrum given in Eq.~(\ref{arlev}). Firstly,
the partition function is automatically convergent. Secondly, the grand canonical 
approach is compatible with Hawking radiation in the sense that it requires the 
density operator to vary with time. Thirdly, the mean square
fluctuations of the energy comes out positive. 
The distribution of the area (energy) levels is of
the Gaussian type (for $\langle {\bf A}\rangle\gg {\cal L}_{p}^{2}$), with
relative fluctuations $\Delta {\bf A}/\langle {\bf A}\rangle$ of order
of ${\cal L}_{p}/\sqrt{\langle {\bf A}\rangle}$. Furthermore, as a byproduct of 
the grand canonical approach we have recovered the same logarithmic correction 
to the entropy earlier derived by Kastrup~\cite{Kast97}
and by M\"{a}kel\"{a} and Repo~\cite{MakRep98}. Other 
authors~\cite{MaSo97,So94,Car95} have also obtained a logarithmic 
corrections to the entropy of BHs.

\section*{Acknowledgments}
I would like to thank Prof.~Jacob Bekenstein 
for his guidance and support during the course of 
this work. It is also a pleasure to thank A.~E.~Mayo for a lot of help and 
S.~Hod and L.~Sriramkumar for helpful discussions. 
This research was supported by a grant from the Israel 
Science Foundation, established by the Israel National 
Academy of Sciences.

\end{document}